\documentclass{article}
\usepackage{emulateapj,apjfonts}
\input psfig

\def\Sec{\hbox{${}^{\prime\prime}$\llap{.}}}

\def\mh{M_{\bullet}}

\lefthead{Ferrarese \& Merritt}
\righthead{A Fundamental Relation Between Supermassive Black Holes 
and Their Host Galaxies}

\begin{document}

\title{A Fundamental Relation Between Supermassive Black Holes 
and Their Host Galaxies}

\author{Laura Ferrarese\altaffilmark{1}}
\affil{University of California, Los Angeles, CA, 90095, 
laura@astro.ucla.edu}
\authoraddr{Department of Physics and Astronomy, 8371 Math Sciences 
Building, Box 951562, Los Angeles CA 
90095-1562, USA}
\author{David Merritt}  
\affil{Rutgers University, New Brunswick, NJ, 08854, 
merritt@physics.rutgers.edu}
\authoraddr{Department of Physics and Astronomy, 136 Frelinghuysen Road, 
Piscataway, NJ 08854}

\altaffiltext{1}{Current Address: Department of Physics and Astronomy, 
Rutgers University, New Brunswick, NJ 08854}

\begin{abstract}

The masses of supermassive black holes correlate almost perfectly with
the velocity dispersions of their host bulges,
$\mh\propto\sigma^{\alpha}$,  where $\alpha=4.8\pm 0.5$.  The
relation is much tighter than the relation between $\mh$ and bulge
luminosity, with a scatter no larger than expected on the basis of
measurement error alone. Black hole masses estimated by Magorrian  et
al. (1998) lie systematically above the $\mh-\sigma$ relation  defined
by more accurate mass estimates, some by as much as two orders of
magnitude. The tightness of the  $\mh-\sigma$ relation implies a
strong link between black hole formation and the properties of the
stellar bulge.

\end{abstract}

\section{Introduction}

After decades of indirect and circumstantial evidence, the motion of
gas and stars on parsec scales has provided irrefutable dynamical
evidence for the presence of $10^7-10^9$ M$_{\odot}$ black holes (BHs)
in about a dozen elliptical and a handful of spiral galaxies (Kormendy
\& Richstone 1995).  While efforts to build a larger, statistically
significant sample continue, we have now moved from debating the
existence of supermassive BHs, to asking what regulates their
formation and evolution, and how their presence influences, and is
influenced by, their host galaxies.

In an early review based on eight detections, Kormendy \& Richstone
(1995) found that BH masses $\mh$ scale linearly with the absolute
blue luminosity of the host bulge or elliptical galaxy.  This
correlation was later strengthened by Magorrian et al.  (1998) using a
larger ($\sim 30$) sample of galaxies to which simple stellar
dynamical models were applied. At the same time, it has been noted
(e.g.   Jaffe 1999) that the $\mh-B_T^0$ relation suffers from
observational biases and exhibits a large scatter which is  not
accounted for by the uncertainties in the individual measurements.

By understanding how the properties of BHs relate to those of their
host galaxies,  we can hope to learn about the formation and evolution
of both. In this Letter, the connection between BH masses and the
stellar  velocity dispersion of the host galaxy is investigated for
the first  time.   We find a remarkably tight correlation with
negligible intrinsic scatter when using galaxies with well-determined
BH masses (roughly speaking,  those galaxies in which the observations
have resolved the sphere of  gravitational influence of the BH).   Our
results suggest that the stellar  velocity dispersion may be the
fundamental parameter regulating the evolution of supermassive  BHs in
galaxies.

\section{Database}

All secure BH mass estimates available to date (see \S 3), together
with a compilation of properties of the host galaxies, are given in
Table 1. Revised Hubble and T-type (from the Third Reference
Catalogue, RC3, de Vaucouleurs et al.  1991) are found in columns 2
and 3, while column 4 lists distances to the host galaxy.  With a few
exceptions detailed in the footnotes, all distances are from surface
brightness fluctuation (SBF) data  (Tonry et al. 2000) calibrated as
in  Ferrarese et al.  (2000).

Total apparent magnitudes m$_{B}$, uncorrected for Galactic
absorption, are from the RC3 for all elliptical galaxies (T-type =
$-$4 or smaller), and from de Vaucouleurs \& Pence (1978) for the
Milky Way.  For the lenticular and spiral galaxies (T-type = $-$3 and
larger), $m_{B}$ for the bulge is derived using the empirical
correlation between T-type and the ratio between bulge and total
luminosity (Simien \& de Vaucouleurs 1986), and is deemed to be
accurate within 0.5 mag.  Finally, all  magnitudes are corrected for
Galactic extinction using the DIRBE/IRAS  maps of Schlegel, Finkbeiner
\& Davis  (1998) and an  extinction law following Cardelli, Clayton \&
Mathis (1989), and  converted to absolute magnitudes (col. 5) given
the distances  in column 4.

The methods used in deriving the BH masses, and references to the
original papers are listed in the last column of Table 1.   Because
the masses depend linearly on the assumed distance to the host
galaxies, the values in column 6 have been corrected to adhere to our
homogeneous set of distances.  This correction is random in nature,
and negligible with the exception of IC 1459, which is twice as
distant as assumed in the original paper.  Uncertainties in the host
galaxies' distances  have been incorporated in the errors in the BH
masses.

Elliptical galaxies and bulges of spirals have radial velocity
gradients, hence a measure of the velocity dispersion $\sigma$ will
depend on the  distance to the galaxy, the size of the aperture used,
and the location of the aperture with respect to the  galaxy core
(e.g.  Davies et al.  1987). For this work, we have chosen the same
definition of $\sigma$ used for studies of the fundamental plane of
elliptical galaxies, namely the {\it central} velocity dispersion,
typically  measured in an aperture a few arcseconds in diameter
(Davies et al. 1987, and  references listed in the footnotes of Table
1). Our choice will be justified in \S 3.  To bring all values of
$\sigma$ to a common system, we have adopted the prescription of
Jorgensen, Franx \& Kjaergaard (1995) and transformed all velocity
dispersions to the equivalent of an aperture of radius $r_e/8$, where
$r_e$ is the galaxy (or bulge) effective radius.  The applied
corrections are very small (rarely exceeding 5\%) and are deemed
accurate to within 1\% (Jorgensen, Franx \& Kjaergaard 1995).  Raw and
corrected $\sigma$ are listed in columns 7 and 8 of

\hskip -0.7in\psfig{file=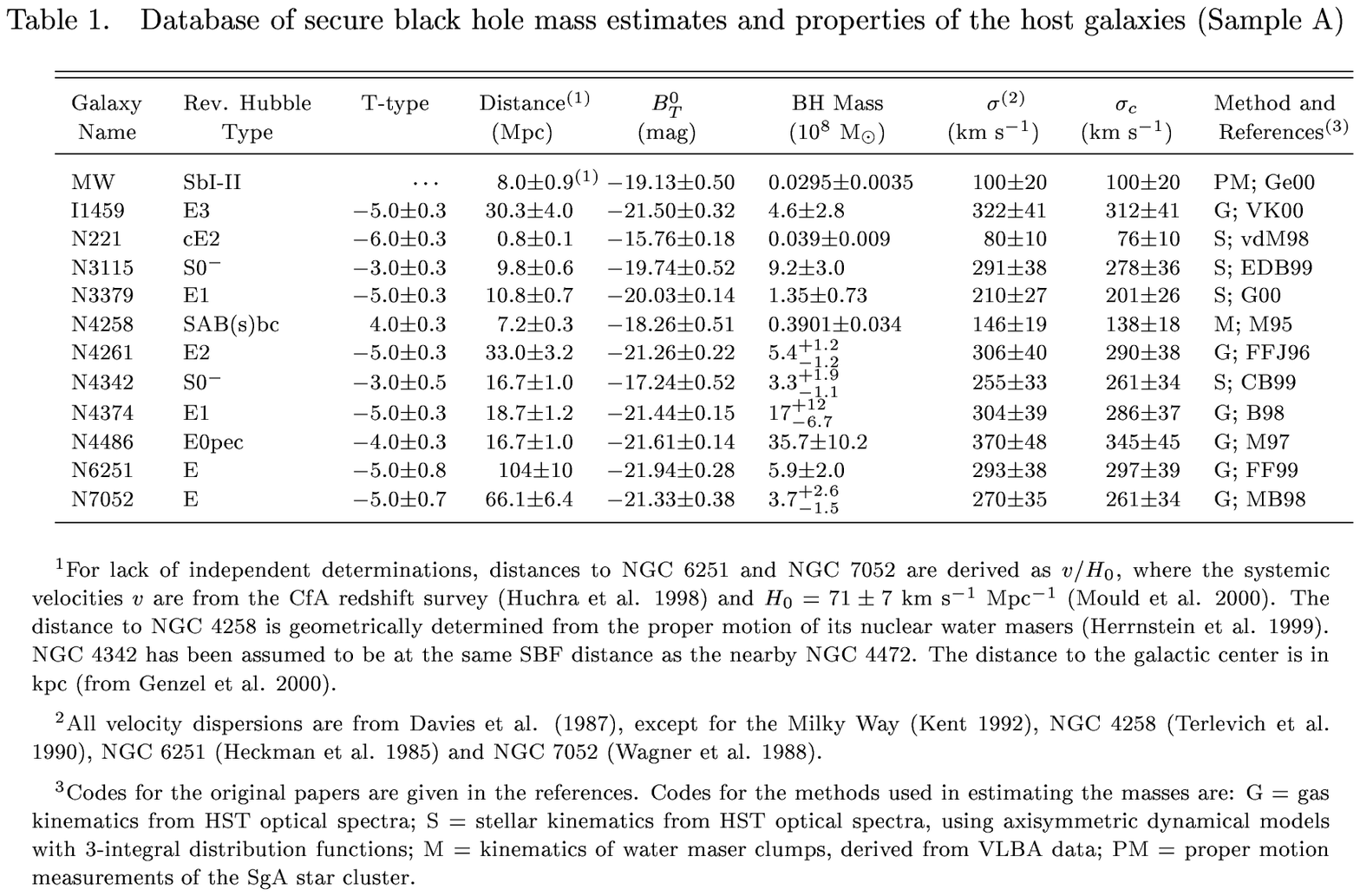,width=17.5cm,angle=0}
\vskip -4.3in

\noindent Table 1. For the Milky Way, we adopt the mean velocity
dispersion compiled by Kent (1992) within $r_e/8$.

\section{Analysis}

We did not make any attempt to homogenize the error estimates on the
BH masses.  Except for adding in quadrature the (small) uncertainty in
the galaxies' distances, the errors on $\mh$ listed in Table 1 are
those quoted by the respective authors.  However, the real
uncertainties are often much larger.  For instance, Magorrian et al.
(1998) derived BH masses based on fitting a simple  class of dynamical
models to ground-based kinematical data.  In almost  all of these
galaxies, the data can equally well be fit by a  more general class of
model with no BH at all; thus  the Magorrian et al.  mass estimates
might conservatively be  interpreted as upper limits (e.g. van der
Marel 1997).  The same is true  for the majority of ground-based,
stellar-kinematical BH detections.

In view of this fact, we list in Table 1 only the galaxies which we
deem to have reliable BH mass estimates.  The proper motion studies of
the Sagittarius A star cluster and the dynamics of the water maser
disk in NGC 4258 lead to the most robust determinations of $\mh$.
Close seconds are estimates in 10 additional galaxies, based on data
from high-resolution HST observations, either absorption-line stellar
spectra or observations of the motion of nuclear dust/gas disks.
These 12 galaxies comprise our ``Sample A.''  Additional galaxies with
less secure BH mass detections are  Arp 102B ($\mh$ obtained from
fitting accretion disk models to variable optical emission lines,
Newman et al. 1997), and all galaxies for which $\mh$ is estimated
based on stellar kinematics obtained from the ground  (e.g. the
Magorrian et al.  1998 sample). These galaxies define our ``Sample
B'': the data are tabulated in Merritt \& Ferrarese (2000).

\hskip 5in\psfig{file=tab1.ps,width=17.5cm,angle=0}
\vskip -4.3in

We then searched for linear correlations between $\log\mh$ and  both
$B_T^0$ and $\log \sigma_c$.   We used the bivariate linear regression
routine of  Akritas \& Bershady (1996), which accommodates intrinsic
scatter as  well as measurement errors in both variables.  $\mh$ was
taken as  the dependent variable.  The results of the regression fits,
applied to each sample of $N$ galaxies, are summarized in Table 2 and
Figure 1.

The correlation between $\mh$ and bulge magnitude $B_T^0$  (Fig.
1a,c) is poor, both for Sample A and Sample B. Although the  best
linear fit to the data has a slope close to the value of $-0.4$
expected if $\mh$ is simply proportional to the bulge mass, it is
apparent from the figure, and from the reduced $\chi^2_r$ of the fit
(Table 2), that even by restricting the sample to the galaxies with
the  most accurately determined BH masses, the intrinsic scatter in
the  $\mh - B_T^0$ relation remains significantly larger than the
reported  errors.  No sub-sample of galaxies, selected either by
Hubble Type or  method used in deriving $\mh$, defines a tight linear
relation  between $\mh$ and $B_T^0$. This implies that differences in
the  mass-to-light ratio between Hubble types, or systematic biases
affecting any particular method, are unlikely to account for the large
scatter.
 
Figures 1b and 1d show the dependence of $\mh$ on the central  stellar
velocity dispersion $\sigma_{c}$ of the host bulge or  elliptical
galaxy.  The correlation is remarkable: Sample A, which  shows a large
scatter in the $\mh - B_T^0$ plots, now defines a  linear relation
with negligible intrinsic scatter.  The best-fit  linear relation is
\begin{equation}
\log\mh = 4.80 (\pm 0.54) \log \sigma_c - 2.9 (\pm 1.3)
\end{equation}
with $\mh$ in units of $M_{\odot}$ and $\sigma_c$ in km s$^{-1}$.  The
slope of the relation remains unaltered, albeit with a larger
uncertainty, if the two galaxies at the low-velocity-dispersion  end
of the distribution (the Milky Way and M32) are excluded from the fit.
The reduced $\chi^2$ of the fit (Table 2) is only $0.8$, consistent
with a scatter that derives entirely from  measurement errors. The
first incarnation of Eq. (1) was suggested by Merritt (2000) (the
``Faber-Jackson law for black holes'').

The galaxies in Sample B define a much weaker correlation between
$\mh$ and $\sigma_c$ (Fig.  1d).  Furthermore, the BH  masses in this
sample lie systematically above the mean line defined  by Sample A,
some by factors of $\sim 10^2$. Two  factors distinguish the two
samples: the reliability of the  $\mh$ estimates; and the method used
to derive the BH masses.  About 1/2 of the mass determinations in
Sample A are based on gas motions while almost all of the Sample B
masses are derived from stellar kinematics. We see no evidence for a
systematic difference between the two types of mass determination; for
instance, NGC 4342 and NGC 7052 have identical  $\sigma_c$  and $\mh$,
even though the determination of $\mh$ in NGC 4342 is based on stellar
kinematics and in NGC 7052 on rotation of a gas disk.   In the case of
IC 1459, for which $\mh$ predicted by  Eq. 1 is $2.5$ times larger
than measured, Verdoes-Kleijn et al. (2000) suggest that the true BH
mass could be  a factor $3-4$ greater than their best estimate due to
non-circular motions of the gas.  It seems likely that the different
correlations defined by the two samples result largely from errors in
the determination of $\mh$  for the galaxies in Sample B.

Our choice of aperture-corrected, central velocity dispersions is
convenient but not unique. We note first  that correcting $\sigma$ for
the effect of aperture size does not introduce a bias in either the
slope or the intercept (see Table 1). However, the need for

\vskip 0.2in

\psfig{file=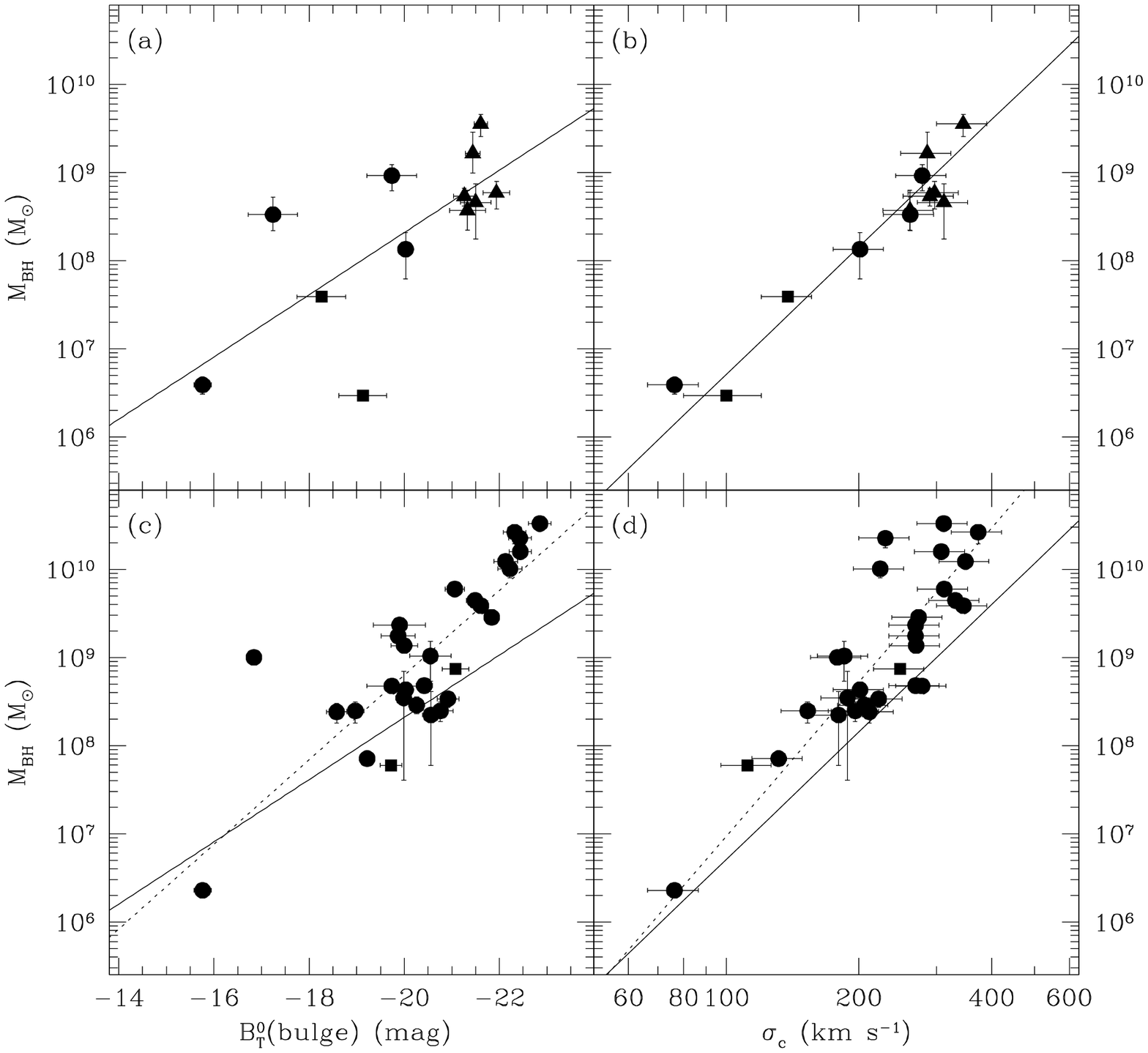,width=8.8cm,angle=0} \figcaption[fig1.ps]{(a): BH
mass versus absolute  blue luminosity of the host elliptical galaxy or
bulge for our most  reliable Sample A.  The solid line is the best
linear fit (Table 2).   Circles and triangles represent mass
measurements  from stellar and dust/gas disk kinematics respectively.
The squares are  the Milky Way ($\mh$ determined from stellar proper
motions) and NGC 4258 ($\mh$ based on water maser kinematics), the
only two spiral galaxies in  the sample.  (b) Again for Sample A, BH
mass versus the central  velocity dispersion of the host elliptical
galaxy or bulge, corrected  for the effect of varying aperture size as
described in \S 2.   Symbols are as in panel (a).  (c): Same as panel
(a) but for Sample B.  Circles are elliptical galaxies, squares are
spiral galaxies.   The solid line is the same least-squares fit shown
in panel (a); the dashed line is the fit to Sample B.  All BH mass
estimates in this sample are based on stellar kinematics.   (d): Same
as panel (b) but for Sample B. Symbols are as in panel (c).}
\hskip -.2in\psfig{file=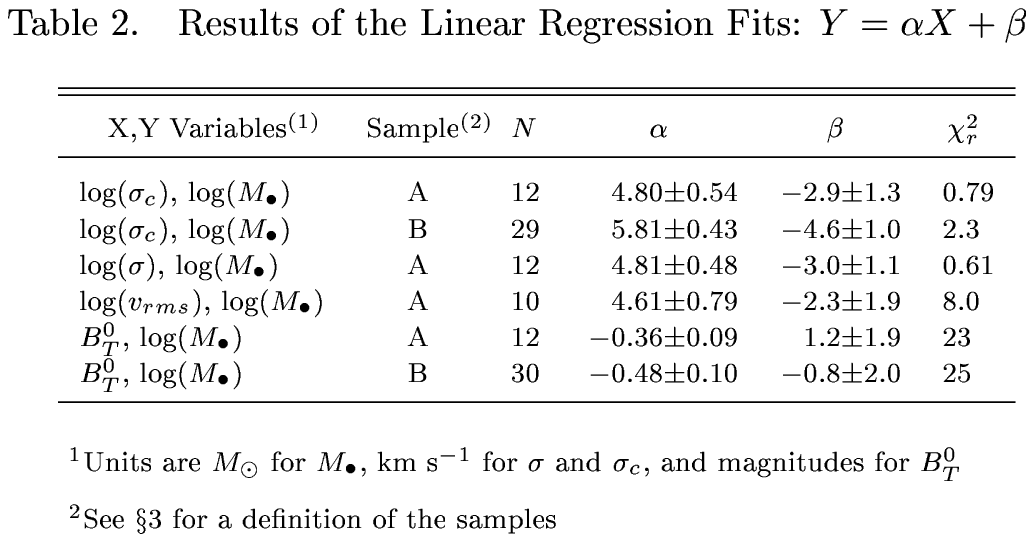,width=17.5cm,angle=0}
\vskip -6.7in

\noindent aperture corrections could be avoided by using a measurement
of the rms velocity at some fiducial distance from the center.  Figure
2 plots $\mh$ vs.  the rms  stellar velocity $v_{rms}$ at $r_e/4$,
with $v_{rms} = \sqrt{(\sigma^{2} + v_{r}^{2}/\sin^2
i)_{r_{e}/4}}$. Here $\sigma$ and $v_{r}$ are the measured stellar
velocity dispersion and mean line-of-sight velocity respectively.  A
complication with this approach is the typically poorly-constrained
value of the inclination angle $i$ between the rotation axis and the
line of sight. Estimates for $i$ are available only for NGC 3115
(Emsellem, Dejonghe \& Bacon 1999) and NGC 4342 (Scorza \& van den
Bosch 1998).  Neglecting or wrongly estimating  $\sin i$  will
increase the scatter in the relation and bias  the slope too low, by
moving faint, rapidly-rotating galaxies to the left in the
$\mh-v_{rms}$ plane. Nevertheless, for our sample A, linear regression
fits (Table 2)  show that the slopes of $\mh$ vs bulge velocity are
coincident whether $\sigma_c$ or $v_{rms}$ is used.

An interesting question is whether the tight correlation between $\mh$
and $\sigma_c$ might simply reflect the influence of the BH on the
stellar kinematics of the nucleus.  The coincidence of the slopes
obtained when $v_{rms}$ is substituted for $\sigma_c$ is the most
convincing evidence that this is not the case, since  $v_{rms}$ is
measured well beyond the radius at which the BH could have a
measurable effect.  In addition, most of the measurements of $\sigma$
listed in Table 1 were  carried out using apertures much larger than
the expected radius of  gravitational influence of the BH.  We stress
that -- even if the correlation between $\mh$ and $\sigma_c$ were due
in part to the gravitational influence of the BH on the motion of
stars in the nucleus -- this would not vitiate the usefulness of the
relation as a predictor of $\mh$.  {\it Figure 1b suggests  that $\mh$
can be predicted with an accuracy of $\sim 30\%$ or better  from a
{\it single}, low-resolution observation of a galaxy's velocity
dispersion.}  This is a remarkable result.

\section{Discussion}

We have found a nearly perfect correlation between the masses of
nuclear BHs and the velocity dispersions of their host bulges,
$\mh \propto \sigma^{\alpha}$, $\alpha= 4.8\pm 0.5$.  Here we
examine some of the implications of this correlation.

The Magorrian et al. (1998) mass estimates fall systematically above
the  tight correlation defined by our Sample A (Fig. 1d), some by as
much as two orders of magnitude. The discrepancy is a strong function
of  distance to the galaxy, particularly at the high-mass end: nine of
the  Magorrian et al. galaxies have BH masses that are larger than the
largest BH mass in our Sample A ($3.6\times 10^9 M_{\odot}$, in NGC
4486),  and six of these are more distant than 50 Mpc.  A number of
authors (van der Marel 1997; Ho 1998)  have suggested on other grounds
that the Magorrian et al. mass estimates may be systematically
high. If our Eq. (1) correctly predicts $\mh$, the gravitational
radius of influence of the BHs in most of these galaxies  would be far
too small to have been resolved from the ground.  For example, Eq. (1)
predicts $\mh \sim 2.8 \times 10^8 M_{\odot}$ for  NGC 4874, a full
two orders of magnitudes smaller than the Magorrian et al.  estimate;
the implied radius of influence is $\sim 24$ pc $\sim$ 0\Sec05.  In
support of this idea, we note that the best-fitting $\mh$ found by
Magorrian et al. in five of their 36 galaxies was negative, while an
additional three galaxies -- altogether, $1/4$ of their sample -- were
consistent with  $\mh<0$.  In view of this, we suggest that
correlation studies based  on the Magorrian et al. masses
(e.g. Merrifield, Forbes \& Terlevich 2000)  be interpreted with
caution.

\psfig{file=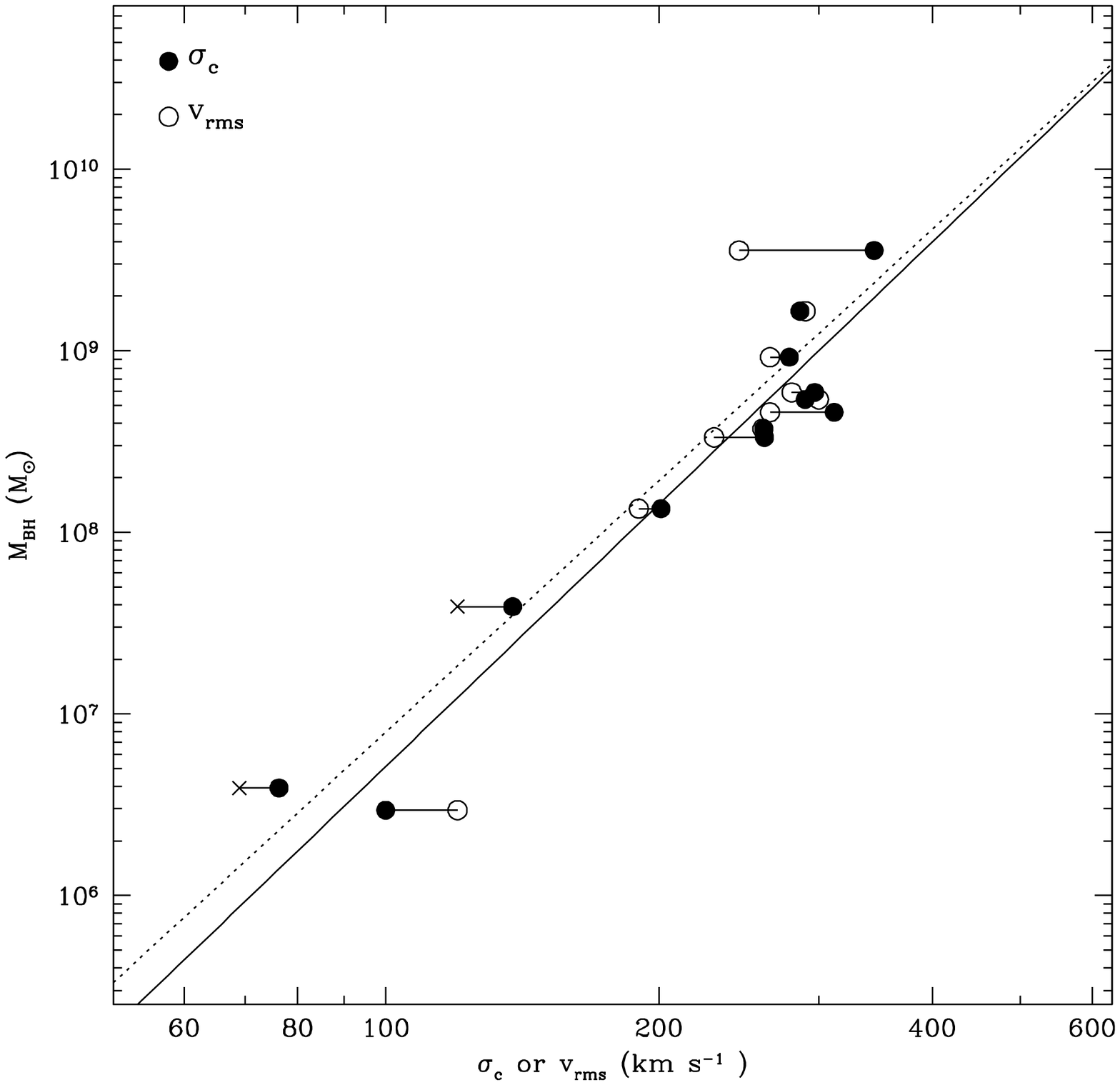,width=8.8cm,angle=0}  \figcaption[fig2.ps]{BH mass
versus the central velocity dispersion $\sigma_c$ of the host
elliptical galaxy or bulge (solid circles)  or the rms velocity
$v_{rms}$ measured at 1/4 of the  effective radius (open circles).
Crosses represent lower limits in $v_{rms}$.  The solid and dashed
lines are the best linear fits using  $\sigma_c$ (as in Figure 1b) and
$v_{rms}$ respectively.}
\vskip .3in

In passing, we caution against the indiscriminate extrapolation of
Eq. (1) much below the range plotted in Figure 1 (for example to the
range appropriate to dwarf elliptical galaxies or globular clusters),
as the formation mechanism of BHs with masses smaller than $\sim
10^{5}$ M$_{\odot}$ might differ from that of more massive systems
(Haehnelt, Natarajan \& Rees 1998).

Why should BH masses be so tightly correlated with bulge velocity
dispersions?  One possibility is a fundamental connection between
$\mh$ and bulge mass, with $\sigma$ a good predictor of bulge mass --
a better predictor, for instance, than $B_T^0$.  This explanation is
superficially plausible,  since the masses of early-type galaxies
scale with their luminosities as $M\sim L^{5/4}$ (Faber et al. 1987)
and $L\sim \sigma^4$,  hence $M\sim\sigma^5$.  The $\mh-\sigma$
relation of Figure 1b would therefore imply a rough proportionality
between BH mass and bulge mass, i.e. that a universal fraction of the
baryonic mass was converted into BHs.  However  early-type galaxies
appear to be two-parameter systems (Djorgovski \& Davis 1987) and it
is not clear that $\sigma$ alone should be a good  predictor of galaxy
mass.

Another possibility is that $\sigma$ measures the depth of the
potential well in which the BH formed.  A number of authors  (Silk \&
Rees 1998; Haehnelt, Natarajan \& Rees 1998) have suggested that
quasar outflows might limit BH masses by inhibiting accretion of gas.
Equating the energy liberated in one dynamical time of the bulge to
the gravitational binding energy, and assuming accretion at the
Eddington rate, gives a maximum BH mass that scales as $\sigma^5$
(Silk \& Rees 1998),  again consistent with the observed relation.
This dependence could be maintained in the face of mergers only if BHs
continued to grow by gas accretion during all stages of the merger
hierarchy (Kauffmann \& Haehnelt 2000).

\vskip .1in
We thank Avi Loeb for an illuminating discussion which served as a key
motivation for this paper, and W. Dehnen, C. Joseph, J. Magorrian and
M. Sarzi for  comments and suggestions which improved the manuscript.
LF acknowledges grant  NASA NAG5-8693, and DM acknowledges grants
NSF AST 96-17088 and NASA NAG5-6037.  This research has made use of
the NASA/IPAC Extragalactic Database (NED).

\end{document}